\documentclass[superscriptaddress,aps,twocolumn]{revtex4-1}
\usepackage{graphicx}
\usepackage{dcolumn}
\usepackage{bm,bbm}
\usepackage{xcolor}
\usepackage{tcolorbox}
\usepackage{algorithm}
\usepackage{algpseudocode}
\usepackage{amsmath}
\usepackage[toc,page]{appendix}
\usepackage[normalem]{ulem}
\usepackage{makecell}
\usepackage{multirow}

\usepackage[colorlinks,breaklinks]{hyperref}
\makeatletter
\newcommand\org@hypertarget{}
\let\org@hypertarget\hypertarget
\renewcommand\hypertarget[2]{%
  \Hy@raisedlink{\org@hypertarget{#1}{}}#2%
  }
\makeatother
\newcommand{\ket}[1]{\left\vert#1\right\rangle}

\newcommand{\ketbra}[2]{| #1\rangle \langle #2|}
  
\newcommand{\mpi}[1]{\textcolor{olive}{#1}}
\begin{document}

\title{Pathways for entanglement based quantum communication in the face of high noise}

\author{Xiao-Min Hu}
\affiliation{CAS Key Laboratory of Quantum Information, University of Science and Technology of China, Hefei, 230026, People's Republic of China}
\affiliation{CAS Center For Excellence in Quantum Information and Quantum Physics, University of Science and Technology of China, Hefei, 230026, People's Republic of China}
\author{Chao Zhang}
\affiliation{CAS Key Laboratory of Quantum Information, University of Science and Technology of China, Hefei, 230026, People's Republic of China}
\affiliation{CAS Center For Excellence in Quantum Information and Quantum Physics, University of Science and Technology of China, Hefei, 230026, People's Republic of China}
\author{Yu Guo}
\affiliation{CAS Key Laboratory of Quantum Information, University of Science and Technology of China, Hefei, 230026, People's Republic of China}
\affiliation{CAS Center For Excellence in Quantum Information and Quantum Physics, University of Science and Technology of China, Hefei, 230026, People's Republic of China}
\author{Fang-Xiang Wang}
\affiliation{CAS Key Laboratory of Quantum Information, University of Science and Technology of China, Hefei, 230026, People's Republic of China}
\affiliation{CAS Center For Excellence in Quantum Information and Quantum Physics, University of Science and Technology of China, Hefei, 230026, People's Republic of China}
\author{Wen-Bo Xing}
\affiliation{CAS Key Laboratory of Quantum Information, University of Science and Technology of China, Hefei, 230026, People's Republic of China}
\affiliation{CAS Center For Excellence in Quantum Information and Quantum Physics, University of Science and Technology of China, Hefei, 230026, People's Republic of China}
\author{Cen-Xiao Huang}
\affiliation{CAS Key Laboratory of Quantum Information, University of Science and Technology of China, Hefei, 230026, People's Republic of China}
\affiliation{CAS Center For Excellence in Quantum Information and Quantum Physics, University of Science and Technology of China, Hefei, 230026, People's Republic of China}
\author{Bi-Heng Liu}
\email{bhliu@ustc.edu.cn}
\affiliation{CAS Key Laboratory of Quantum Information, University of Science and Technology of China, Hefei, 230026, People's Republic of China}
\affiliation{CAS Center For Excellence in Quantum Information and Quantum Physics, University of Science and Technology of China, Hefei, 230026, People's Republic of China}
\author{Yun-Feng Huang}
\affiliation{CAS Key Laboratory of Quantum Information, University of Science and Technology of China, Hefei, 230026, People's Republic of China}
\affiliation{CAS Center For Excellence in Quantum Information and Quantum Physics, University of Science and Technology of China, Hefei, 230026, People's Republic of China}
\author{Chuan-Feng Li}
\email{cfli@ustc.edu.cn}
\affiliation{CAS Key Laboratory of Quantum Information, University of Science and Technology of China, Hefei, 230026, People's Republic of China}
\affiliation{CAS Center For Excellence in Quantum Information and Quantum Physics, University of Science and Technology of China, Hefei, 230026, People's Republic of China}
\author{Guang-Can Guo}
\affiliation{CAS Key Laboratory of Quantum Information, University of Science and Technology of China, Hefei, 230026, People's Republic of China}
\affiliation{CAS Center For Excellence in Quantum Information and Quantum Physics, University of Science and Technology of China, Hefei, 230026, People's Republic of China}
\author{Xiaoqin Gao}
\email{xgao5@uottawa.ca}
\affiliation{Institute for Quantum Optics and Quantum Information (IQOQI), Austrian Academy of Sciences, Boltzmanngasse 3, 1090 Vienna, Austria}
\affiliation{Vienna Center for Quantum Science and Technology (VCQ), Faculty of Physics, University of Vienna, Boltzmanngasse 5, 1090 Vienna, Austria}
\affiliation{Department of physics, University of Ottawa, Advanced Research Complex, 25 Templeton Street, K1N 6N5, Ottawa, ON, Canada}
\author{Matej Pivoluska}
\email{pivoluskamatej@gmail.com}
\affiliation{Institute of Computer Science, Masaryk University, 602 00 Brno, Czech Republic}%
\affiliation{Institute of Physics, Slovak Academy of Sciences, 845 11 Bratislava, Slovakia}%
\author{Marcus Huber}
\email{marcus.huber@univie.ac.at}
    \affiliation{Vienna Center for Quantum Science and Technology, Atominstitut, TU Wien,  1020 Vienna, Austria}%
\affiliation{Institute for Quantum Optics and Quantum Information (IQOQI), Austrian Academy of Sciences, Boltzmanngasse 3, 1090 Vienna, Austria}

\begin{abstract}
Entanglement based quantum communication offers an increased level of security in practical secret shared key distribution. One of the fundamental principles enabling this security -- the fact that interfering with one photon will destroy entanglement and thus be detectable -- is also the greatest obstacle. Random encounters of traveling photons, losses and technical imperfections make noise an inevitable part of any quantum communication scheme, severely limiting distance, key rate and environmental conditions in which QKD can be employed. Using photons entangled in their spatial degree of freedom, we show that the increased noise resistance of high-dimensional entanglement, can indeed be harnessed for practical key distribution schemes. We perform quantum key distribution in eight entangled paths at various levels of environmental noise and show key rates that, even after error correction and privacy amplification, still exceed $1$ bit per photon pair and furthermore certify a secure key at noise levels that would prohibit comparable qubit based schemes from working.
\end{abstract}
\date{\today}
\maketitle

Quantum key distribution (QKD) \cite{BENNETT20147,E91, RevModPhys.81.1301,Lo2014,2019arXiv190601645P} is one of the most prominent and mature applications of quantum information theory. 
It can be used to establish a shared and private random bit-string among two parties, that can subsequently be used to encrypt information \cite{Delfs2015}. 
There are different levels of security of quantum key distribution, depending on the assumptions placed on each of the devices used. 
The weakest form are the so-called prepare and measure protocols \cite{BENNETT20147,6state,PhysRevLett.68.3121}, which assume a trusted source of quantum states in possession of one of the parties, as well as perfectly characterised measurement devices for both parties. 
Although such assumptions about components of QKD implementation are often reasonable, they open up loopholes which the potential adversary can abuse to perform attacks on the implementation of the protocol \cite{lydersen2010hacking,qian2018hacking}.
The other extreme is given by so-called device independent quantum key distribution \cite{Pironio_2009,PhysRevLett.113.140501,10.1145/2885493,EAT,Murta19}, where no assumptions are placed on any devices, except for the privacy of locally generated randomness. 
Such protocols provide a revolutionary paradigm shift in designing secure QKD protocols, but they remain largely impractical, because they require a loophole-free Bell inequality violations, which can be obtained only in strict laboratory conditions \cite{hensen2015loophole,shalm2015strong,giustina2015significant}.
In-between these two extremal cases, there are plenty of scenarios with various levels of trust placed on the devices, which leads to very different practically achievable key rates \cite{PhysRevLett.108.130503,PhysRevA.84.010302,PhysRevA.85.010301,PhysRevA.86.062319,PhysRevLett.111.130502,PhysRevLett.117.190501,curty2014finite,Pirandola2015}. 
Entanglement based protocols belong to this last group as they typically assume the entanglement source is in the control of the adversary.
This makes entanglement protocols secure against many attacks abusing source imperfections (e.g. photon splitting attack \cite{brassard2000limitations,lutkenhaus2002quantum}), possible against prepare and measure protocols.

The physical principle ensuring security of quantum key distribution protocols can be intuitively understood from two fundamental facts about of quantum physics. 
First of all, an unknown quantum state cannot be copied (no-cloning theorem \cite{Park1970,Wootters1982,DIEKS1982271}) and second, a state cannot undergo a measurement procedure without being influenced (projection postulate of quantum mechanics, see e.g. \cite{10.5555/1972505}). 
So when encoding information in individual quantum systems, it is impossible to intercept and learn information from them, without also revealing one's presence. 
While this principle enables classically unachievable levels of security, it also presents a serious challenge. 
Any interaction of these individual quantum systems with an environment, any background photons that are accidentally detected and other imperfections in the devices will manifest as noise in the data. 
Such environmental noise cannot be distinguished from noise that would result from malicious activity.
There are two big challenges of contemporary QKD stemming from noise \cite{Diamanti2016}.
First, QKD protocols cannot certify any shared key, if the noise level is above certain threshold.
Second, environmental noise significantly affects the achievable key rate of many protocols even in relatively low noise regimes.
One of the big remaining challenges of QKD is therefore to design protocols, which can tolerate large amounts of environmental noise and produce large amounts of key in moderate noise regimes.

The potential way to solve both of these challenges by employing high-dimensional degrees of freedom of photons (see \cite{doi:10.1002/qute.201900038}) has been proposed as early as $20$ years ago \cite{PhysRevA.61.062308,PhysRevLett.88.127902}.
The idea of increased key rate is straightforward -- one photon carrying information in $d$ dimensional degree of freedom (called a \emph{qudit}) can produce as much as $\log_2 d$ bits of randomness.
Simultaneously, in theory, increasing the dimension $d$ of used quantum systems also increases the amount of tolerable noise \cite{PhysRevA.82.030301}.
Practical demonstrations of high-dimensional QKD (HDQKD) followed much later.
Prepare and measure protocols demonstrated that in low noise regimes one can indeed obtain increased key-rates \cite{Etcheverry2013,Mirhosseini_2015,PhysRevA.96.022317,islam2017provably,Sit:17,Ding2017,PhysRevApplied.11.064058,Islam_2019,PhysRevApplied.14.014051}. 
On the other hand entanglement based HDQKD protocols were achieved only by employing additional assumptions about the distributed state \cite{Zhong_2015}, 
thus compromising the source independence of the protocol, or restricted measurements \cite{Gr_blacher_2006,PhysRevA.88.032305}.
Additionally, none of the implementations show exceptionally high noise resistance. 
This is partially caused by the fact that with increasing the dimensions in real experiment, one inevitably also increases the environmental noise (see \cite{doda2020quantum}). Further, this increased noise takes an extra toll, as error correction requires more communication in higher dimensions.

In this article we present a first experimental demonstration of an entanglement based HDQKD protocol, which does not impose any assumptions about the distributed state. 
This is possible thanks to several recent breakthroughs.
It was recently shown \cite{PhysRevX.9.041042} that high-dimensional entanglement, i.e. entanglement in multiple degrees of freedom can exhibit an increased resistance to real physical noise compared to low-dimensional counterparts. 
This led to the proposal of a QKD protocol, simultaneously coding in multiple subspaces of high-dimensional states \cite{doda2020quantum} \textcolor{black}{(see also \cite{PhysRevA.98.062301})}, theoretically predicting the possibility of establishing a secure key in the presence of unprecedented noise levels. 
The last recent breakthrough is the development of experimental setups for creation and manipulation of path entanglement \cite{hu2020efficient}, which allow implementation of true multi-outcome measurements with high fidelity.
Putting these ideas together, we implement the protocol introduced in \cite{doda2020quantum} using eight-dimensional path entanglement and bilateral eight-outcome measurements. 
We show that even after post-processing, the key rate exceeds one bit per coincidence, i.e. each detected pair establishes more key than would be possible to encode in even a perfect and noiseless qubit. 
Furthermore, we prepare an entire family of states by adding artificial noise to the experiment, fully exploring the achievable noise resistance of the protocol. 

\begin{figure*}[tph!]
\includegraphics [width= 1\textwidth]{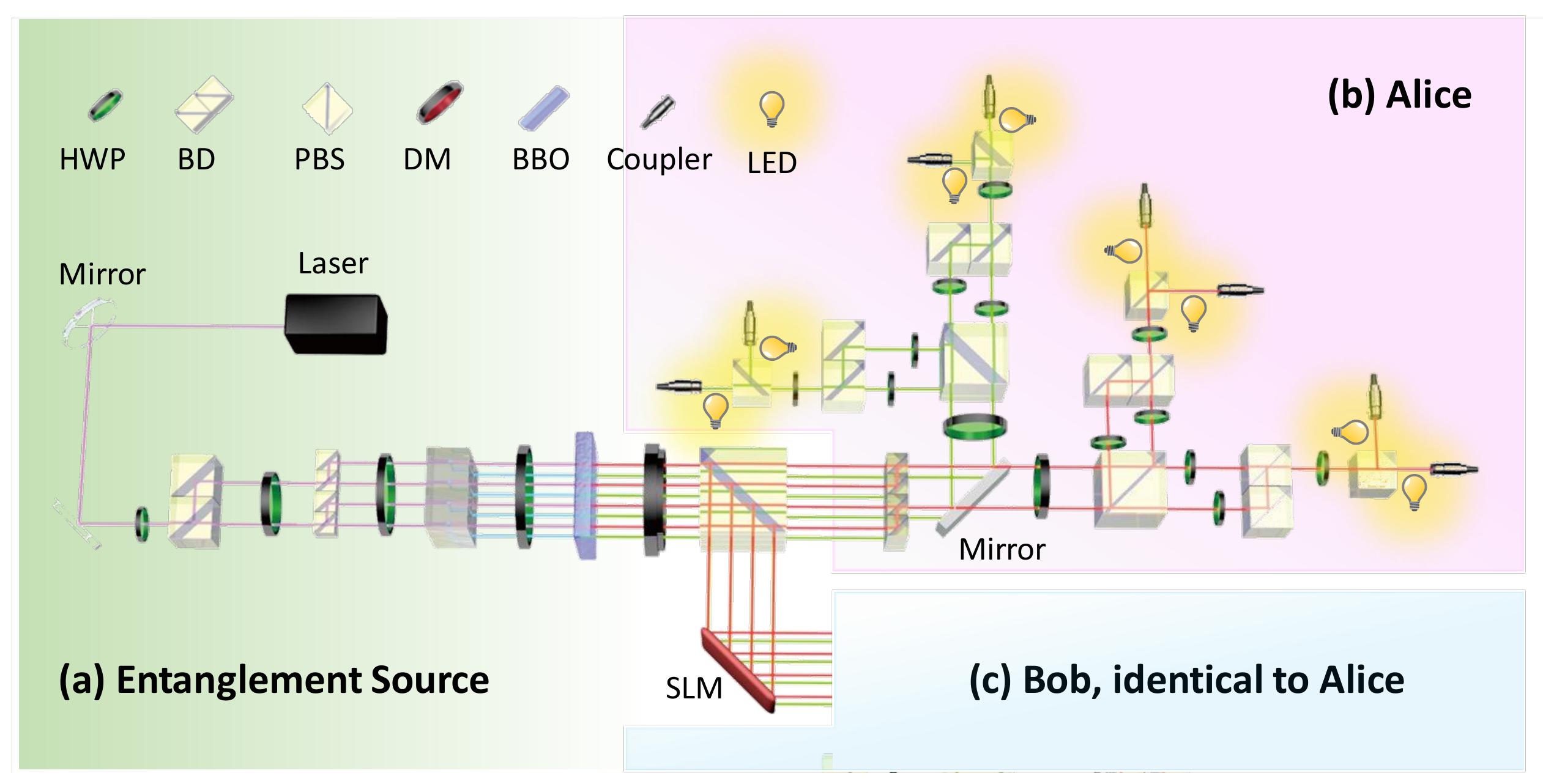}
\vspace{-0.3cm}%
\caption{Experimental setup. (a) Preparation of eight-dimensional entanglement: Eight parallel beams are obtained by eight equal divisions of the continuous-wave light ($@404 \ nm$, the diameter is $0.6 \ mm$), with the help of three half-wave plates (HWPs $@22.5^{\circ}$) and three beam-displacers (BDs). Eight beams are assigned into two-layer eight paths, the upper layer and the lower layer represented with purple and blue colours respectively, \textcolor{black}{labelled with "0", "1", \dots, "7"}. The distance between two neighboring paths is $2 \ mm$. Another HWP $@22.5^{\circ}$ is necessary for all beams to transmit with $H$ polarization. \textcolor{black}{Each beam injects into a BBO crystal will generate infrared photon pairs called single and idler photons ($H^{classic}_{404nm} \xrightarrow{SPDC}\ket{H}_{808nm}\otimes\ket{V}_{808nm})$ via the Type-II spontaneous parametric down-conversion (SPDC)~\cite{schneeloch2019introduction,ou2007multi} (see the supplementary materials). 
Due to the eight beams pumping the BBO crystal being coherent, the photon pairs are generated in a coherent superposition in different paths. 
Hence an eight-dimensional path-based entangled target state $|\phi^{+}_8\rangle=1/\sqrt{8}\sum^{7}_{i=0}|ii\rangle$ ($@808 \ nm$), distributed in red layer and green layer respectively, is prepared.} When using only the upper layer, we get a four-dimensional \textcolor{black}{target state $|\phi^{+}_4\rangle=1/\sqrt{4}\sum^{3}_{i=0}|ii\rangle$.} Similarly working with two paths \textcolor{black}{("0" and "1")} only we get a two-dimensional \textcolor{black}{target state $|\phi^{+}_2\rangle=1/\sqrt{2}\sum^{1}_{i=0}|ii\rangle$.}
\textcolor{black}{The remaining 404 $nm$ beams are removed after the BBO crystal by a dichroic mirror (DM) and the photons pairs are separated by a polarizing beam splitter (PBS).} $H$ photons are sent to Alice, while $V$  photons are sent to Bob after using a phase-only spatial light modulator (SLM) to manipulate the phase of incident photons. 
Parts (b) and (c) of the figure depict multi-outcome measurements for Alice and Bob. {Sixteen adjustable intensity LED light sources in front of 16 couplers are used to introduce noise on each detector.} The conversion of projective measurements between computational basis and subspace Fourier-transform basis can be realized by changing the angles of HWPs, (see the Appendix).
}
\label{fig:subspace}
\vspace{0.2cm}%
\end{figure*}
First, let us briefly review the high-dimensional entanglement based QKD protocol developed in \cite{doda2020quantum}. 
The protocol is composed of $N$ rounds, in which the source distributes a $d\times d$ entangled state $\rho_{AB}\in \mathcal{H}_A\otimes\mathcal{H}_B$ to two communicating parties, Alice and Bob.
In the ideal case $\rho_{AB} = \ketbra{\phi^+_d}{\phi^+_d}$, where $\ket{\phi^+_d} = \frac{1}{\sqrt{d}}\sum_{i=0}^{d-1} \ket{ii}$.  
Postulates of quantum mechanics guarantee that measuring this state by both Alice and Bob in the $d$-dimensional computational basis (called a \emph{key basis}) leads to two $\log_2(d)N$-bit strings $X$ and $Y$. These two strings are uniformly distributed, perfectly correlated and private, thus they constitute a shared secret key.
However, any real world implementation is necessarily imperfect and thus the quality of the state $\rho_{AB}$ needs to be assessed.
Particularly, in randomly chosen rounds, Alice and Bob measure the state in a \emph{test basis} which allows them to estimate the amount of key they can distill from their key basis measurement outcomes $X$ and $Y$.
This step is then followed by classical post-processing. This is composed of \emph{error correction}, in which differences between $X$ and $Y$ are corrected and \emph{privacy amplification}, in which the final key -- shorter but uniformly distributed shared string -- is obtained.
We employ methods developed in \mpi{\cite{PhysRevA.82.030301}}, where the quality of data obtained in the $d$-dimensional key basis is assessed by measurements in a mutually unbiased basis \cite{doi:10.1063/1.2713445}.
The test rounds of the protocol are then used to assess the following
{error vector:
\begin{align}\label{eq:witness}
   \vec{e}_{\mathrm{t}} = \left(e_{\mathrm{t}}^{(0)},e_{\mathrm{t}}^{(1)},\dots,e_{\mathrm{t}}^{(d-1)}\right),
\end{align}
where $e_\mathrm{t}^{(j)} = \sum_{i=0}^{d-1}\Pr(i,i+j \mod d \vert\mathrm{test})$ 
is the probability that Alice obtained result $i$ and Bob obtained result $i+j \mod d$, when they were both measuring in their test basis.}
{This error vector is used to bound the adversary's information in the asymptotic regime against collective attacks  as $H(\vec{e}_\mathrm{t})$, where $H(\cdot)$ is the Shannon entropy function \cite{PhysRevA.82.030301}. 
A similar error vector $\vec{e}_\mathrm{k}$ can be defined for the measurements in the key basis. 
In turn, $H(\vec{e}_\mathrm{k})$ gives the amount of information Alice and Bob need to exchange in the error correction phase.
Together, the asymptotic key rate per coincidence of a $d$-dimensional instance of the described entanglement based protocol can be estimated as 
\begin{equation}
    K_d \geq \log_2(d) - H(\vec{e}_\mathrm{k})
    -H(\vec{e}_\mathrm{t}).\label{eq:keyRate}
\end{equation}}

A key technique we use from \cite{doda2020quantum} is the splitting of the $d$-dimensional Hilbert space into $d/k$ mutually exclusive subspaces of size $k$ leading to additional post-selection -- Alice and Bob keep the measurement outcomes, only if they obtained results in the same subspace. 
The key rate is obtained separately in each subspace and the final key rate is obtained as an average of $d/k$ observed key rates.
 
\begin{figure*}[!tbh]
\includegraphics [width= 0.98 \columnwidth]{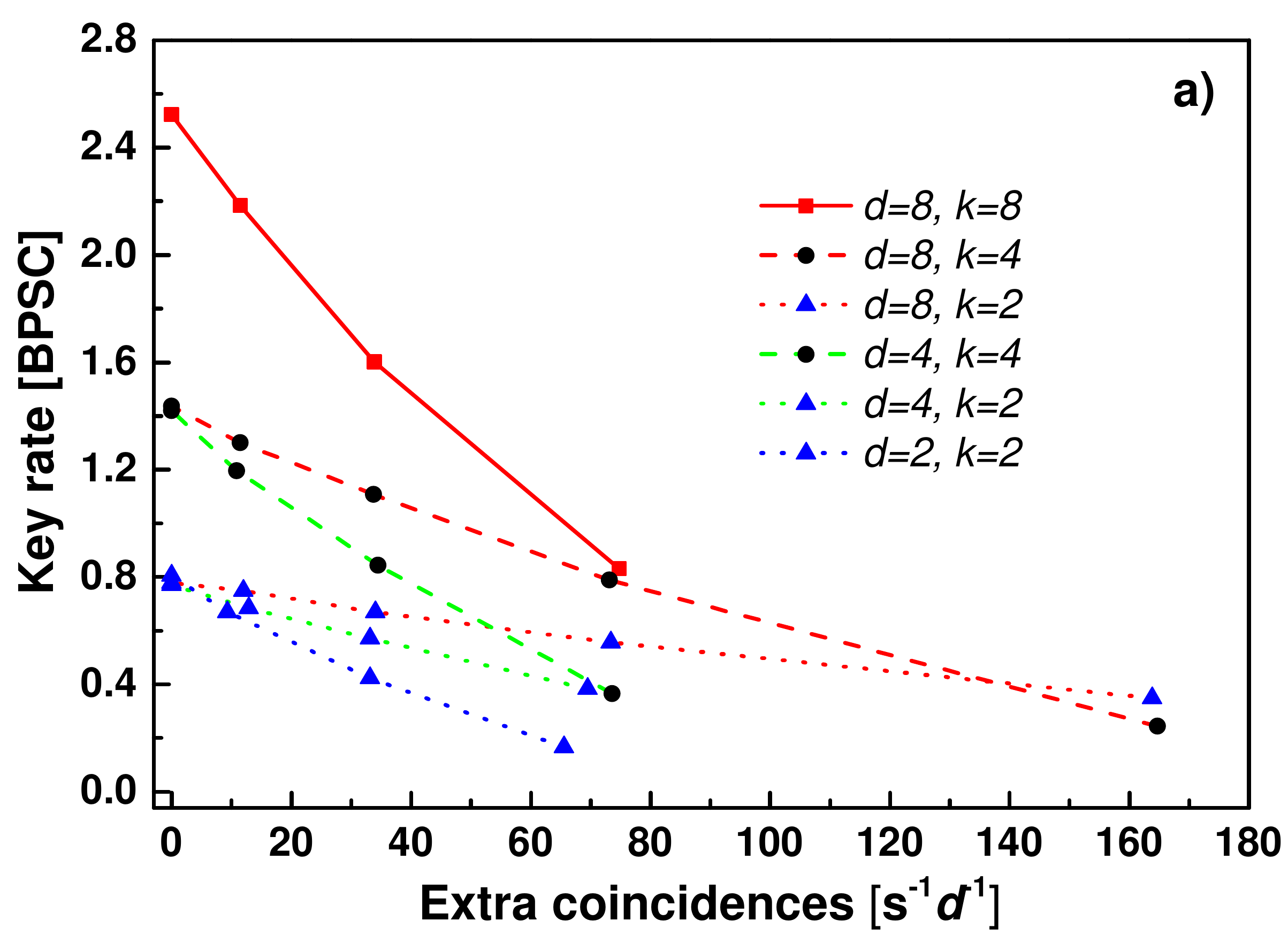}
\includegraphics [width=\columnwidth]{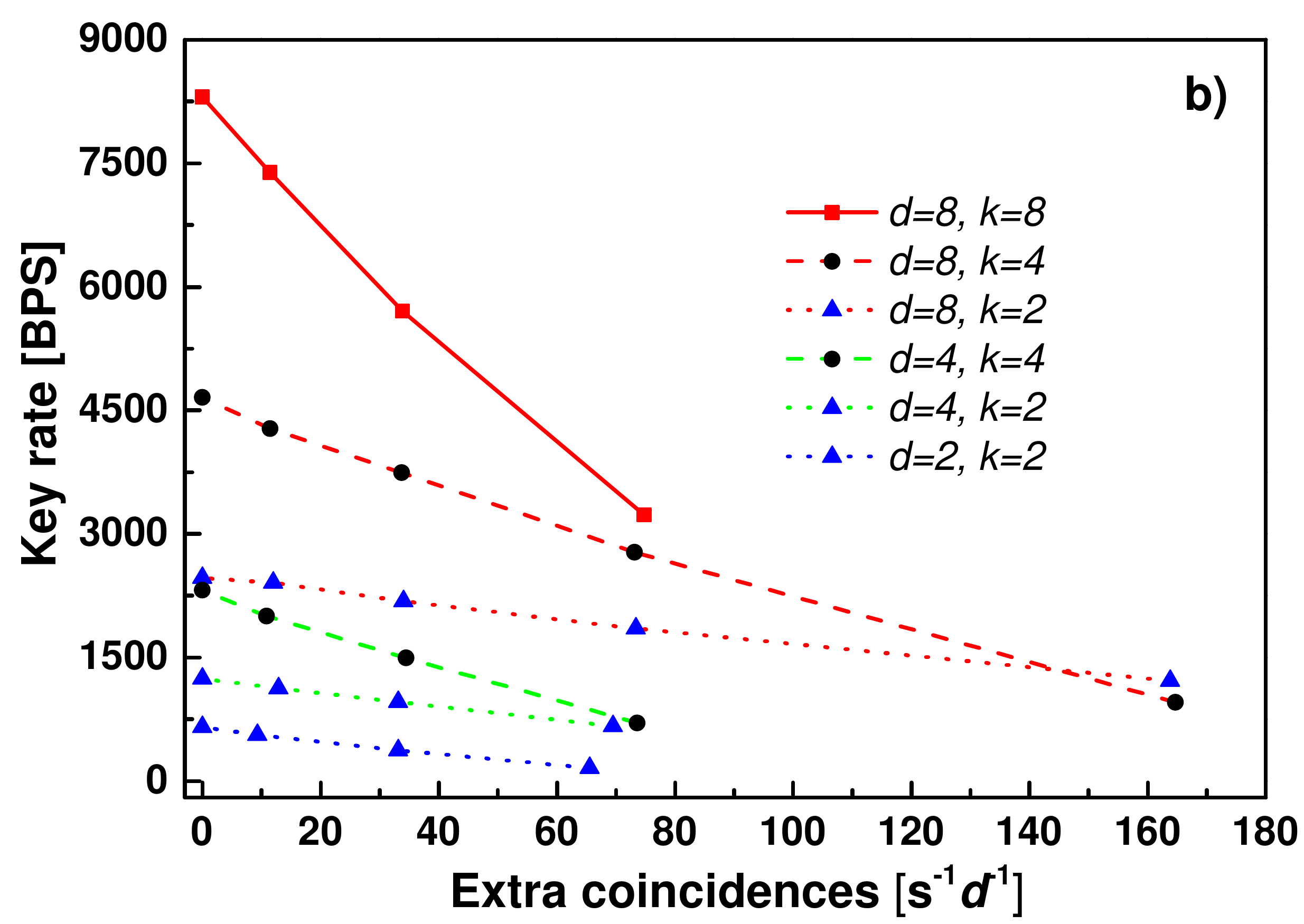}
\vspace{-0.2cm}%
\caption{\textcolor{black}{The key rate of bits per subspace post-selected coincidence (BPSC, a)) and of bits per second (BPS, b)) obtained in the eight/four/two (red/green/blue) - dimensional spaces.}  
Noise is shown as average additional coincidences per second, divided by the local dimension ($8$, $4$ and $2$ respectively).
The points {(error bars are inside the symbols)} represent the experimental values obtained by adding different levels of noise. 
The accurate experimental data is shown in {Tables S6-S11} of the Appendix. 
}
\label{fig:keyrate}
\vspace{0.2cm}%
\end{figure*}

In our experiment, we thus aim at creating a maximally entangled state in all $d$ dimensions using path entanglement. To fully explore the high-noise regime in a controlled manner, we shine ambient light on each detector. 

To explore the interplay of global and subspace dimensions, we study three cases of  global dimension $d=8$, $d=4$ and $d=2$, with subspace dimensions $k=2,\ 4$ and $8$.
For preparing the eight-dimensional target state $|\phi^{+}_8\rangle$, encoded in the path degree of freedom, we
use three half-wave plates (HWPs) $@22.5^{\circ}$ together with three beam-dispalcers (BDs). Eight parallel beams are distributed to eight paths with the same energy by dividing \textcolor{black}{the pump light equally}. The light is produced by a continuous-wave diode laser $@404 \ nm$, as shown in figure \ref{fig:subspace} (a).
Remarkably, it is easy to prepare the four-dimensional target state $|\phi^{+}_4\rangle$ if we only consider the upper layer (marked red in figure \ref{fig:subspace}).
To compensate for the phase between Alice and Bob, a spatial light modulator (SLM) is added to implement an arbitrary phase on the vertically polarised light \cite{hu2020efficient}. 

In our setup we use polarization to control the path degree of freedom in order to implement $8$-outcome measurements required for the protocol. Note, however, that in principle, our multi-outcome measurement technique can be generalized to higher dimensions effectively \cite{hu2020efficient}. By changing the angles of HWPs placed in parts (b) and (c) of figure \ref{fig:subspace}, Alice and Bob can switch between the projective measurements used in the protocol (see the Appendix). 
Due to current limitations on the parallelism of beams in the BD, the mutually unbiased basis in dimension $8$ would not reach the desired fidelity, a fact that will in the future be mitigated by improvements in BD manufacturing. 
Nonetheless, we generalised the protocol to work with mutually unbiased subspace measurements with overlapping subspaces to certify security in $8$ dimensions, even without fully mutually unbiased measurements as described in the original protocol (see the Appendix).
We record coincidences between all paths and compute both the secure key generated per selected photon pair (i.e. the average key rate per subspace post-selected coincidence $K_{BPSC}$) and the resulting secure key per second ($
K_{BPS} = K_{BPSC} \times TSCS,$
where $TSCS$ is the total subspace coincidence per second). These results are plotted in figure \ref{fig:keyrate}a) and \ref{fig:keyrate}b). The key rate is computed from raw data by following the {subspace} protocol from Ref.~\cite{doda2020quantum} {with key rate formula presented in Eq. \eqref{eq:keyRate}} for different levels of physical noise, i.e varying environmental conditions created by adding physical noise to the setup in a controlled fashion. This is achieved by putting independent noise sources in front of each optical coupler to introduce white noise, as shown in figure \ref{fig:subspace} (b) and (c) respectively. These extra sources of noise lead to accidental coincidences in the data, which we use as a measure of physical environmental noise in the setup. {Our noise parameter is described by the number of accidental coincidences added to the measurement data per second divided by the local dimension $d$, but it can be equivalently expressed by the value of parameter $p$ in the experimental state 
\begin{equation}\label{eq.Noisefraction}
    \rho_d = (1-p)\rho_d^\mathrm{ent} + p\frac{I_{d^2}}{d^2},
\end{equation}
where $\frac{I_{d^2}}{d^2}$ is the completely mixed state of a $d\times d$ dimensional quantum state, $\rho^\mathrm{ent}_d$ is the actual state our entanglement setup produces and $p\in\{0,0.025,0.075,0.15,0.3\}$} {(see the Appendix).}

We perform six separate experiments, for $8$, $4$ and $2$ local paths (i.e. local dimensions) and subspace measurements in dimensions $2$ and $4$.

We observe that for low noise, we can obtain much higher key rate $K_{BPSC}$ by setting the subspace dimension $k$ higher for the same global dimension $d$. 
However, with  the noise increasing, using the subspaces with lower dimensions leads to stronger noise-robustness.
From the experimental results, we can see the key rate $K_{BPSC}$ of $k=4$ decreases rather fast compared to the cases with $k=2$.
Similar results are shown in subspaces with different dimensions when $d=4$.
Importantly, the robustness of the protocol also increases with the total dimension $d$. 
For example, the key rate $K_{BPSC}$ of $k=4$ decreases more slowly in $d=8$ than in $d=4$ and similar observation can be made for $k=2$.

{
One can notice that for all subspace sizes, $K_{BPS}$ is effectively doubled when one compares  $d=8$ to $d=4$ and $d=2$. This occurs because doubling $d$ also doubles the number of entangled pairs generated per second, as more beam paths are collected in detectors. $TSCS$ therefore increases from
$\approx 800$ pairs per second in case of $d=2$ to
$\approx 1600$ pairs per second in case of $d=4$ and $\approx 3200$ pairs per second in case of $d=8$ (see the Appendix). 
Note, however, that the increase of $TSCS$ for higher dimensions can be expected also for fundamental reasons. 
Considering the damage threshold of nonlinear crystals (such as BBO) \cite{huang2011experimental}, the permitted maximal pump strength is proportional to the path dimension $d$ and one can, in principle, create more entangled pairs for higher $d$.
This is because in path entanglement the crystal is pumped at multiple distinct locations and therefore heated more evenly.}

The intricate relation between global and subspace dimension shows a clear pathway towards optimal usage of high-dimensional entanglement for quantum communication. 
While increasing the global dimension improves the achievable key rate and noise resistance simultaneously, it should be noted that it of course comes at the cost of increasing the number of detectors on each side. 
Another interesting factor is the optimal subspace dimension, as it clearly shows that for low noise levels a high subspace dimension is optimal.
On the other hand the noise resistance is achievable with decreasing subspace dimension as a function of noise. 
In experimental setups with constant signal to noise ratios this implies a single optimal subspace coding.
In variable situations, such as complex quantum networks or free space communication it would seem prudent to consider an on-the-fly optimisation of subspace dimension to swiftly adapt to changing conditions.
The particular scheme we use for creating spatial entanglement carries another distinct advantage for quantum communication. 
The fact that we coherently split the beam prior to pumping the crystal means that the pump laser is heating the crystal in a more distributed fashion, allowing for larger crystals and larger pump intensities before a limiting intensity is reached. 
This increases the potential number of entangled  pairs per second and carriers with it the potential to again increase the key rate by another physical mechanism.
{We also want to point out that there remains one significant pathway to improve the key rate by implementing more than two MUB measurements in the test rounds. As shown in \cite{PhysRevA.82.030301} using multiple MUB measurements should lead to increase in both amount of certified bits per round and noise resistance. 
However, we expect that high total dimension $d$ and subspace size $k=2$ will lead to the greatest noise resistance even in protocols using multiple MUB measurements. This is based on the following intuition: In high noise regimes the distributed entangled state $\rho$ can be expected to have Schmidt number equal to $2$ and thus measurements in subspaces of size $2$ are best suited to fully utilize it in a QKD protocol.}

In conclusion, by implementing the first entanglement based, high-dimensional and multi-outcome QKD experiment, we were able to achieve key rates exceeding $1$ bit of perfect key after error correction per photon pair. This significant increase even survived the artificial injection of additional accidentals through ambient light. 
By increasing the artificial noise, we were also able to demonstrate the superior noise resistance of subspace coding in high-dimensional systems and experimentally explore the intricate relationship between global dimension, subspace dimension, key rate and noise. 
Our experiment proves the viability of high-dimensional coding for overcoming some of the most significant challenges of quantum communication and identifies novel pathways for noise resistant key distribution.
\textcolor{black}{Phase-stable distribution of path-based entangled states in real experimental conditions is a significant challenge that needs to be addressed before our approach can be used in practice. 
Path to OAM conversion \cite{Fickler2014} or multicore fibres \cite{DaLio2021,hu2020efficientdistribution} could be the missing ingredient to take this proof of principle demonstration towards practical QKD implementation.} {Finally, the improved rate of entanglement distribution will be of interest to other entanglement based applications beyond QKD.}

\section*{References}
\bibliography{bibliography}

\section*{Acknowledgements}
This work was supported by the National Key Research and Development Program of China (No.\ 2017YFA0304100, No. 2016YFA0301300 and No. 2016YFA0301700), National Natural Science  Foundation of China (Nos. 11774335, 11734015, 11874345, 11821404, 11904357), the Key Research Program of Frontier Sciences, CAS (No.\ QYZDY-SSW-SLH003), Science Foundation of the CAS (ZDRW-XH-2019-1), the Fundamental Research Funds for the Central Universities, USTC Tang Scholarship, Science and Technological Fund of Anhui Province for Outstanding Youth (2008085J02), Anhui Initiative in Quantum Information Technologies (Nos.\ AHY020100, AHY060300). X.G. acknowledges the support of Austrian Academy of Sciences (ÖAW) and Joint Centre for Extreme Photonics (JCEP). M.H. acknowledges funding from the Austrian Science Fund (FWF) through the STARTproject Y879-N27. M.P. acknowledges the support of VEGA project 2/0136/19 and GAMU project MUNI/G/1596/2019.

\clearpage
\onecolumngrid
\appendix

\section{Calculating key rate in $d=8,k=8$}\label{d8keyrate}
While for $d=8$, direct implementation of the Fourier basis with $k=8$ was too noisy in our implementation, we developed another method for certifying security with $8$-outcome measurements. 
By measuring projections onto different global bases $(\ket{a}\pm\ket{b})\otimes (\ket{a}\pm\ket{b})$ and $(\ket{a}\pm i \ket{b}) \otimes (\ket{a}\pm i \ket{b})$ for each pair of modes $a<b$ (i.e. $56$ different measurement settings), we are able to directly lower bound the fidelity $F_+$ of the experimental state $\rho_8$ to the maximally entangled state $|\phi^{+}_8\rangle$. This method was developed in \cite{hu2020efficient}. 
Recall that in order to estimate the information of the adversary about Alice's measurement results in the computational basis, we need to {lower bound the entropy of the error vector 
\begin{equation}
     \vec{e}_{\mathrm{t}} = \left(e_{\mathrm{t}}^{(0)},e_{\mathrm{t}}^{(1)},\dots,e_{\mathrm{t}}^{(d-1)}\right),
\end{equation}
where $e_{\mathrm{t}}^{(0)}$ is the probability that Alice and Bob obtained a perfectly correlated measurement outcomes in the test basis.
Note that $e_{\mathrm{t}}^{(0)}$ can be written as
\begin{equation}\label{eq:witness}
    e_{\mathrm{t}}^{(0)}  = \sum_{i=0}^7 \langle \widetilde{i}\widetilde{i}^*|\rho_8| \widetilde{i}\widetilde{i}^*\rangle,
\end{equation}} 
where $|\widetilde{i}\rangle$ is a basis mutually unbiased to the computational basis in $d=8$ case. Here we utilize the fact that $\rho$ can be decomposed as $\rho_8 = F_+(|\phi^{+}\rangle\langle\phi^{+}|) + (1-F_+)\rho^{\perp}_8$, therefore 
\begin{align}
    {e_{\mathrm{t}}^{(0)}} &= \sum_{i=0}^7 \langle\widetilde{i}\widetilde{i}^*|\rho_8| \widetilde{i}\widetilde{i}^*\rangle\nonumber\\
    &= \sum_{i=0}^7F_+\langle\widetilde{i}\widetilde{i}^*|\phi^{+}_8\rangle\langle\phi^{+}_8|\widetilde{i}\widetilde{i}^*\rangle + (1-F_+)\langle\widetilde{i}\widetilde{i}^*|\rho^{\perp}_8| \widetilde{i}\widetilde{i}^*\rangle\nonumber \\
    &= F_+ + (1-F_+)\langle\widetilde{i}\widetilde{i}^*|\rho^{\perp}_8| \widetilde{i}\widetilde{i}^*\rangle\nonumber\\
    &\geq F_+,\label{eq:d8LB}
\end{align}
where the last inequality holds because $(1-F_+)\langle\widetilde{i}\widetilde{i}^*|\rho^{\perp}_8| \widetilde{i}\widetilde{i}^*\rangle \geq 0$.
Eq. \eqref{eq:d8LB} can directly be used to {upper bound} {$H(\vec{e}_{\mathrm{t}})$} as
{
\begin{equation}
\label{eq:Pg}
   H(\vec{e}_{\mathrm{t}})\geq H\left(e_{\mathrm{t}}^{(0)},\tfrac{e_{\mathrm{t}}^{(0)}}{d-1},\dots,\tfrac{e_{\mathrm{t}}^{(0)}}{d-1}\right),
\end{equation}
}
{this follows directly from the properties of Shannon entropy, which achieves its maximum for a uniform distribution.}
In order to obtain the desired asymptotic key rate we also need to calculate the amount of information Alice needs to share in order to perform  the error correction {$H(\vec{e}_{\mathrm{k}})$}. This quantity can be directly calculated from the measurement outcomes in the computational basis in the $d=8$ case.

\section{Multi-outcome measurement}
\begin{figure*}[!h]
\includegraphics [width= 1\textwidth]{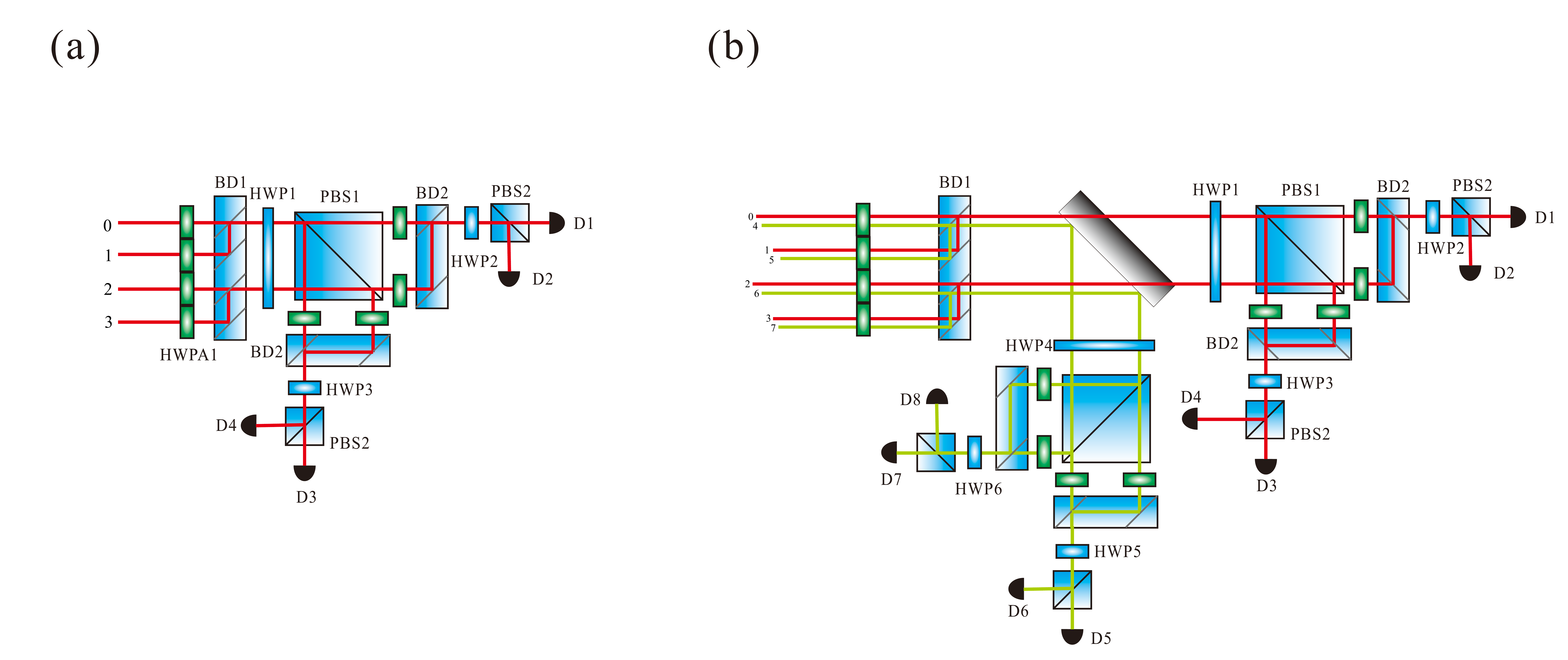}
\caption{Four and eight-outcome measurement setups. 
(a) Four-outcome measurement setup. 
With adjusting the angles of HWPs (1-3), the computational basis and Fourier-transform basis of $k=2,4$ subspace can be realized. For details, see Tables \ref{d4k4} and \ref{d4k2}.
(b) Eight-outcome measurement setup. 
Via adjusting the angles of HWPs (1-6), the computational basis and Fourier-transform basis of $k=2,4$ subspace can be realized. 
Eight beams distribute in upper and lower layers marked with red and green colors respectively.
Only the upper layer is used for the four-outcome measurement setup.
For details, see Tables \ref{d8k4} and \ref{d8k2}.
}
\label{fig:setup}
\end{figure*}

Path entanglement has shown its advantages, such as it is possible to implement non-trivial $d$-outcome measurements. 
Here we introduce how to construct a four-outcome and eight-outcome measurement in our experiment.
We construct the computational basis ($\{|0\rangle, |1\rangle, |2\rangle, |3\rangle\}$) and Fourier basis ($\ket{m_i} = { \frac{1} {\sqrt{4}}} {\sum_{j=0}^{3} \omega_k^{mj}} \left|i_j\right\rangle$, where $\omega_k=e^{i2\pi/k}$.) for the four-dimensional MUB, as shown in figure \ref{fig:setup}(a).
We use a polarizer to control the path degree of freedom, a HWPA1 is used to control the polarizations of \textcolor{black} {paths "0", "1", "2" and "3",} which encode polarizations of $V$, $H$, $V$ and $H$ respectively. 
Via BD1, \textcolor{black}{paths "0", "1" and "2", "3"} are distributed into two polarization-based subspaces respectively. 
Then HWP1 and PBS1 are used to complete the MUB measurement for two polarization-based two-dimensional subspaces. 
The results are used to measure the MUB of each two-dimensional subspace by the sets of $\{$BD2, HWP2, PBS2$\}$ and $\{$BD3, HWP3, PBS3$\}$ respectively.
The four-dimensional MUB measurement can be completed after two levels of two-dimensional subspace coherent measurements. 
Importantly, through the measurement of cascaded two-dimensional subspace, of any $2^{n}$-dimensional MUB measurement can be completed. 
For $d$=2 MUB measurement, we only \textcolor{black} {use "0" and "1" paths}, therefore we only need two detectors D1 and D3. The angles of the HWPs are set in table \ref{d2k2}.
For implementing the projective measurements of the QKD protocol presented in figure \ref{fig:setup} in main text, the angles of HWPs are set in tables \ref{d2k2}, \ref{d4k4}, \ref{d4k2}, \ref{d8k4} and \ref{d8k2}.

Due to the current limitations on the parallelism of beams in BD, the MUB in eight or higher dimensions can not reach the desired fidelity, it can be improved by enhancing the BD technology in future.
Therefore, in our experiment, we only complete $k=2$ and 4-dimensional subspace outcome measurements in the global space $d=8$. 
Instead of measuring the eight-dimensional MUB, we do many measurements in its two-dimensional subspace.
For the case $d = 8, k = 4$, we use a mirror to separate the global space into two four-dimensional subspaces distributed into upper and lower layers. 
Then the $4$-dimensional subspace MUB is obtained.
Notably, the coincidence efficiency of entangled states is roughly $15\%$ without noise. 
With loading noise, the coincidence efficiency will decrease rapidly.

\textcolor{black}{\section{Spontaneous parametric down-conversion process (SPDC) and multiphoton noise}}

\textcolor{black}{Spontaneous parametric down-conversion (SPDC) is a common way to generate correlated photon pairs, historically labeled as signal photon ($s$) and idler photon ($i$). 
In general, the pair of photons are produced by passing the pump light through a nonlinear crystal.  
To generate two-photon entangled states, we use a low gain region $|\eta|<<1$. 
Then, a quantum state can be approximately written as (particle number representation)~\cite{ou2007multi,schneeloch2019introduction}:}
\textcolor{black}{
\begin{equation}
|\Psi\rangle \approx \left(1-|\eta|^{2} / 2\right)|0\rangle+\eta\left|1_{s}, 1_{i}\right\rangle+\eta^{2}\left|2_{s}, 2_{i}\right\rangle+\ldots
\end{equation}}
\textcolor{black}{The first term denotes a pair of photons generated with a probability of $|\eta|^2$ and the second term stands for a quadruplet of photons generated with a probability of $|\eta|^4$. The vacuum state $|0\rangle$ does not directly contribute to photon detection. Due to the imperfect optical elements which lead to the loss of photons and in order to avoid the interference of single photons, we use the standard coincidence counting technique to evidence two photon pairs.
If two photons arrive at Alice's and Bob's detectors in a same time region (coincidence window), we consider them as a pair of photons. If there is only one detector responding in the coincidence window, the result will be discarded.  Of course, single photon counts from detector dark counts or injected environmental photons, as well as multiphoton events will also form coincidences randomly, which will produce noise in the measurement (See "loading noise" section for details). In our experiment, we set the coincidence window with $\tau=5 \ ns$. We can estimate in such way that the probability of each photon pair generated is $|\eta|^{2}=9\times 10^{-5}$. 
Due to $|\eta|<<1$, the probability of generating a quadruplet or more photons is much lower than a pair of photons, therefore
we only consider the first term $\left|1_{s}, 1_{i}\right\rangle$ and can safely ignore the higher-order terms in our experiments.}

\textcolor{black}{We will get photon pairs generated in multiple paths after multiple crystals at the same time if we use coherent pump light. 
Then, a two-photon entanglement is created among different paths, as follows:}
\textcolor{black}{
\begin{equation}
|\Psi\rangle_{entanglement} \approx \eta_{1}\left|1_{s 1}, 1_{i 1}\right\rangle+\eta_{2}\left|1_{s 2}, 1_{i 2}\right\rangle+\eta_{3}\left|1_{s 3}, 1_{i 3}\right\rangle\ldots,
\end{equation}}
\textcolor{black}{where $s1$, $i1$, $s2$, $i2$,$\ldots$ are signal photon and idler photon which can be encoded in any degree of freedom of photons.
In our experiment, we encode the quantum information in the spatial (path) degree of freedom, and we create a high-dimensional entanglement therein. High-dimensional maximally entangled states can be generated by adjusting the probability of each pumped region of the crystal generating a photon pair. } \\

\section{Loading noise}

In our experiment, in order to simulate the noise in the environment, we put sixteen {adjustable intensity LED} light sources before 16 couplers independently to load a certain amount of {single-channel} noise on each detector. 
{The light is coupled to each coupler by diffuse reflection and the total number of photons enter the coupler can be controlled by changing the brightness of the LED light source. According to the brightness of the entanglement source ({coincidence counts}) and the added noise ({coincidence noise}), we then estimate the loaded white noise count on the single channel.}
Assuming the single-channel count of each detector is $S$ {(and thus the combined number of single counts of all detectors on Alice's or Bob's side is equal to $d\times S$, where $d$ is the local dimension)}, the total coincidence counts ($C$) for any two detectors between Alice and Bob is obtained, as follows:{
\begin{equation}\label{eq:extraCOincidences}
C=2\times d\times S\times d\times S \times \tau
\end{equation}
where $\tau$ is the coincidence window.} In our experiment, $\tau=5\times 10^{-9} \ s$ and we can add any proportion of white noise. 

Notably, in the interpretation of results, we calculate the added coincidences from the experimental data -- each $25~s$ run of the protocol is performed at a certain noise level. 
We subtract the total number of coincidences in the noiseless run from each noisy run to obtain extra coincidences. 
Then the additional coincidences per second are divided by the local dimension $d$, which result in the added coincidence counts plotted in figures. 
{There is another way to understand this noise parameter. We are setting the additional noise counts so that the final experimental state has the following form:
\begin{equation}\label{eq:noisefraction}
    \rho_d = (1-p)\rho_d^\mathrm{ent} + p\frac{I_{d^2}}{d^2},
\end{equation}
where $\frac{I_{d^2}}{d^2}$ is the completely mixed state and $\rho_d^\mathrm{ent}$ is the state our entanglement setup produces when the noise LEDs are turned off. In particular, $\rho_d^\mathrm{ent}$ is basically a maximally entangled state in $d$ dimensions, but contains errors due to implementation imperfections (e.g. channel loss, detector inefficiencies and dark counts). Our error therefore simulates a situation when additional light is entering both Alice's and Bob's channel, which is a relevant mode of noise for example in free space implementations. We use $5$ different levels of noise $p\in \{0,0.025,0.075,0.15, 0.3\}$. }

{Note that in a setup with fixed pump strength for all values of $d$, this noise regime corresponds to the situation when a fixed number of photons ({single-channel noise}) enter both channels of Bob and Alice irrespective of the local dimension. 
In particular, intensities of single LEDs (measured in single detector noise counts  $S$ defined above) would need to be halved whenever the dimension $d$ is doubled. 
This would assure that the number of additional coincidences (defined in eq. \ref{eq:extraCOincidences}) stays the same for all values of $d$. 
However, in our implementation we double the strength of pump laser with each doubling of dimension $d$. 
We increase the pump strength in order to showcase that in our particular setup this is physically possible (see conclusions in the main text for discussion and Tables \ref{table:fidelity} - \ref{table:d2k2data} for the number of coincidences per second in different dimensions). 
This increase of pump strength naturally invites a question, how many additional noise coincidences introduced into the experiment would provide a fair comparison of different dimension choices. 
One option is to keep the number of additional photons per party $d\times S$ constant ({single-channel noise}). 
This certainly corresponds to a physical situation, because the amount of additional photons (and therefore coincidences) should not depend on the local dimensions of the setup, but only on outside light conditions. 
However, such noise setup would make the advantage of using high dimensions questionable, because one can expect that the increase of pump strength at constant noise would naturally increase the quality of the data (i.e. signal to noise ratio) irrespective of local dimensions. {To make a fair comparison, we should ensure that the ratio of signal to noise remains constant under different $d$. Therefore, we opted to keep the noise (coincidence noise) fraction $p$ (defined in Eq. \eqref{eq:noisefraction}) constant for all choices of $d$ ({Tables \ref{table:fidelity} - \ref{table:d2k2data}).}} 
In particular this means that the number of added coincidences in our experiment increases with local dimension $d$ (it doubles with each doubling of dimension to compensate for the doubled pump strength). 
The doubling of the number of coincidences corresponds to increasing the number of photons on each side  ({single-channel noise}) by a factor $\sqrt{2}$ with each doubling of the dimensions.
}

\begin{table}[bpth!]
\caption{Measurement for $d=2, k=2$}
\centering
\begin{tabular}{p{5cm}<{\centering}p{5cm}<{\centering}p{5cm}<{\centering}}
\hline
\hline
\multicolumn{3}{c}{\textbf{\textit{Computational basis}}}\\
\hline
HWP1 & HWP2 & HWP3\\
$45^{\circ}$ & $45^{\circ}$ & $45^{\circ}$\\
\end{tabular}

\begin{tabular}{p{7.5cm}<{\centering}p{7.5cm}<{\centering}}
\hline
D1  & D3 \\
$|0\rangle$ & $|2\rangle$  \\
\end{tabular}

\begin{tabular}{p{5cm}<{\centering}p{5cm}<{\centering}p{5cm}<{\centering}}
\hline
\multicolumn{3}{c}{\textbf{\textit{Fourier-transform basis}}}\\
\hline
HWP1 & HWP2 & HWP3\\
$22.5^{\circ}$ & $45^{\circ}$ & $45^{\circ}$\\
\end{tabular}

\begin{tabular}{p{7.5cm}<{\centering}p{7.5cm}<{\centering}}
\hline
D1 & D3 \\
$\frac{1}{\sqrt{2}}(|0\rangle+|1\rangle)$ & $\frac{1}{\sqrt{2}}(|0\rangle-|1\rangle)$ \\
\hline
\hline
\end{tabular}
\label{d2k2}
\end{table}

\begin{table}[bpth!]
\caption{Measurement for $d=4, k=4$}
\centering
\begin{tabular}{p{5cm}<{\centering}p{5cm}<{\centering}p{5cm}<{\centering}}
\hline
\hline
\multicolumn{3}{c}{\textbf{\textit{Computational basis}}}\\
\hline
HWP1 & HWP2 & HWP3\\
$45^{\circ}$ & $45^{\circ}$ & $0^{\circ}$\\
\end{tabular}

\begin{tabular}{p{3.7cm}<{\centering}p{3.7cm}<{\centering}p{3.7cm}<{\centering}p{3.7cm}<{\centering}}
\hline
D1 & D2 & D3 & D4\\
$|0\rangle$ & $|2\rangle$ & $|1\rangle$ & $|3\rangle$ \\
\end{tabular}

\begin{tabular}{p{5cm}<{\centering}p{5cm}<{\centering}p{5cm}<{\centering}}
\hline
\multicolumn{3}{c}{\textbf{\textit{Fourier-transform basis}}}\\
\hline
HWP1 & HWP2 & HWP3\\
$22.5^{\circ}$ & $22.5^{\circ}$ & $22.5^{\circ}$\\
\end{tabular}

\begin{tabular}{p{3.7cm}<{\centering}p{3.7cm}<{\centering}p{3.7cm}<{\centering}p{3.7cm}<{\centering}}
\hline
D1 & D2 & D3 & D4\\
$\frac{1}{2}(|0\rangle+|1\rangle+|2\rangle+|3\rangle)$ & $\frac{1}{2}(|0\rangle+|1\rangle-|2\rangle-|3\rangle)$ & $\frac{1}{2}(|0\rangle-|1\rangle+|2\rangle-|3\rangle)$ & $\frac{1}{2}(|0\rangle-|1\rangle-|2\rangle+|3\rangle)$\\
\hline
\hline
\end{tabular}
\label{d4k4}
\end{table}


\begin{table}[bpth!]
\caption{Measurement for $d=4, k=2$}
\centering
\begin{tabular}{p{5cm}<{\centering}p{5cm}<{\centering}p{5cm}<{\centering}}
\hline
\hline
\multicolumn{3}{c}{\textbf{\textit{Computational basis}}}\\
\hline
HWP1 & HWP2 & HWP3\\
$45^{\circ}$ & $45^{\circ}$ & $45^{\circ}$\\
\end{tabular}

\begin{tabular}{p{3.7cm}<{\centering}p{3.7cm}<{\centering}p{3.7cm}<{\centering}p{3.7cm}<{\centering}}
\hline
D1 & D2 & D3 & D4\\
$|0\rangle$ & $|2\rangle$ & $|1\rangle$ & $|3\rangle$ \\
\end{tabular}

\begin{tabular}{p{5cm}<{\centering}p{5cm}<{\centering}p{5cm}<{\centering}}
\hline
\multicolumn{3}{c}{\textbf{\textit{Fourier-transform basis}}}\\
\hline
HWP1 & HWP2 & HWP3\\
$22.5^{\circ}$ & $45^{\circ}$ & $45^{\circ}$\\
\end{tabular}

\begin{tabular}{p{3.7cm}<{\centering}p{3.7cm}<{\centering}p{3.7cm}<{\centering}p{3.7cm}<{\centering}}
\hline
D1 & D2 & D3 & D4\\
$\frac{1}{\sqrt{2}}(|0\rangle+|1\rangle)$ & $\frac{1}{\sqrt{2}}(|2\rangle+|3\rangle)$ & $\frac{1}{\sqrt{2}}(|0\rangle-|1\rangle)$ &  $\frac{1}{\sqrt{2}}(|2\rangle-|3\rangle)$\\
\hline
\hline
\end{tabular}
\label{d4k2}
\end{table}


\begin{table}[bpth!]
\caption{Measurement for $d=8, k=4$}
\centering
\begin{tabular}{p{2.3cm}<{\centering}p{2.5cm}<{\centering}p{2.5cm}<{\centering}p{2.5cm}<{\centering}p{2.5cm}<{\centering}p{2.3cm}<{\centering}}
\hline
\hline
\multicolumn{6}{c}{\textbf{\textit{Computational basis}}}\\
\hline
HWP1 & HWP2 & HWP3 & HWP4 & HWP5 & HWP6\\
$45^{\circ}$ & $45^{\circ}$ & $45^{\circ}$ & $45^{\circ}$ & $45^{\circ}$ & $45^{\circ}$\\
\end{tabular}

\begin{tabular}{p{1.8cm}<{\centering}p{1.8cm}<{\centering}p{1.8cm}<{\centering}p{1.8cm}<{\centering}p{1.8cm}<{\centering}p{1.8cm}<{\centering}p{1.8cm}<{\centering}p{1.8cm}<{\centering}}
\hline
D1 & D2 & D3 & D4 & D5 & D6 & D7 & D8 \\
$|0\rangle$ & $|2\rangle$ & $|1\rangle$ & $|3\rangle$ & $|4\rangle$ & $|6\rangle$ & $|5\rangle$ & $|7\rangle$\\
\end{tabular}

\begin{tabular}{p{2.3cm}<{\centering}p{2.5cm}<{\centering}p{2.5cm}<{\centering}p{2.5cm}<{\centering}p{2.5cm}<{\centering}p{2.3cm}<{\centering}}
\hline
\multicolumn{6}{c}{\textbf{\textit{Fourier-transform basis}}}\\
\hline
HWP1 & HWP2 & HWP3 & HWP4 & HWP5 & HWP6\\
$22.5^{\circ}$ & $22.5^{\circ}$ & $22.5^{\circ}$ & $22.5^{\circ}$ & $22.5^{\circ}$ & $22.5^{\circ}$\\
\end{tabular}

\begin{tabular}{p{3.7cm}<{\centering}p{3.7cm}<{\centering}p{3.7cm}<{\centering}p{3.7cm}<{\centering}}
\hline
D1 & D2 & D3 & D4\\
$\frac{1}{2}(|0\rangle+|1\rangle+|2\rangle+|3\rangle)$ & $\frac{1}{2}(|0\rangle+|1\rangle-|2\rangle-|3\rangle)$ & $\frac{1}{2}(|0\rangle-|1\rangle+|2\rangle-|3\rangle)$ & $\frac{1}{2}(|0\rangle-|1\rangle-|2\rangle+|3\rangle)$\\
\hline
D5 & D6 & D7 & D8\\
$\frac{1}{2}(|4\rangle+|5\rangle+|6\rangle+|7\rangle)$ & $\frac{1}{2}(|4\rangle+|5\rangle-|6\rangle-|7\rangle)$ & $\frac{1}{2}(|4\rangle-|5\rangle+|6\rangle-|7\rangle)$ & $\frac{1}{2}(|4\rangle-|5\rangle-|6\rangle+|7\rangle)$\\
\hline
\hline
\end{tabular}
\label{d8k4}
\end{table}


\begin{table}[bpth!]
\caption{Measurement for $d=8, k=2$}
\centering
\begin{tabular}{p{2.3cm}<{\centering}p{2.5cm}<{\centering}p{2.5cm}<{\centering}p{2.5cm}<{\centering}p{2.5cm}<{\centering}p{2.3cm}<{\centering}}
\hline
\hline
\multicolumn{6}{c}{\textbf{\textit{Computational basis}}}\\
\hline
HWP1 & HWP2 & HWP3 & HWP4 & HWP5 & HWP6\\
$45^{\circ}$ & $45^{\circ}$ & $45^{\circ}$ & $45^{\circ}$ & $45^{\circ}$ & $45^{\circ}$\\
\end{tabular}

\begin{tabular}{p{1.8cm}<{\centering}p{1.8cm}<{\centering}p{1.8cm}<{\centering}p{1.8cm}<{\centering}p{1.8cm}<{\centering}p{1.8cm}<{\centering}p{1.8cm}<{\centering}p{1.8cm}<{\centering}}
\hline
D1 & D2 & D3 & D4 & D5 & D6 & D7 & D8 \\
$|0\rangle$ & $|2\rangle$ & $|1\rangle$ & $|3\rangle$ & $|4\rangle$ & $|6\rangle$ & $|5\rangle$ & $|7\rangle$\\
\end{tabular}

\begin{tabular}{p{2.3cm}<{\centering}p{2.5cm}<{\centering}p{2.5cm}<{\centering}p{2.5cm}<{\centering}p{2.5cm}<{\centering}p{2.3cm}<{\centering}}
\hline
\multicolumn{6}{c}{\textbf{\textit{Fourier-transform basis}}}\\
\hline
HWP1 & HWP2 & HWP3 & HWP4 & HWP5 & HWP6\\
$22.5^{\circ}$ & $45^{\circ}$ & $45^{\circ}$ & $22.5^{\circ}$ & $45^{\circ}$ & $45^{\circ}$\\
\end{tabular}

\begin{tabular}{p{3.7cm}<{\centering}p{3.7cm}<{\centering}p{3.7cm}<{\centering}p{3.7cm}<{\centering}}
\hline
D1 & D2 & D3 & D4\\
$\frac{1}{\sqrt{2}}(|0\rangle+|1\rangle)$ & $\frac{1}{\sqrt{2}}(|2\rangle+|3\rangle)$ & $\frac{1}{\sqrt{2}}(|0\rangle-|1\rangle)$ & $\frac{1}{\sqrt{2}}(|2\rangle-|3\rangle)$\\
\hline
D5 & D6 & D7 & D8\\
$\frac{1}{\sqrt{2}}(|4\rangle+|5\rangle)$ & $\frac{1}{\sqrt{2}}(|6\rangle+|7\rangle)$ & $\frac{1}{\sqrt{2}}(|4\rangle-|5\rangle)$ & $\frac{1}{\sqrt{2}}(|6\rangle-|7\rangle)$\\
\hline
\hline
\end{tabular}
\label{d8k2}
\end{table}


\section{Experimental results}
Tables \ref{table:fidelity} - \ref{table:d2k2data} contain detailed data represented in figures in the main text. 
NOISE means the average noise coincidences per second divided by the local dimension $d$, thus representing extra coincidences one can assign to each of Alice's detectors. {NOISE corresponds to the noise fraction parameter $p$ as defined in Eq. \eqref{eq:noisefraction}.} BPSC denotes key rate in bits per subspace coincidence, TSCS denotes total number of subspace coincidences per second and BPS denotes key rate in bits per second. {Further, we list error vectors  $\vec{e}_t$ and $\vec{e}_k$ in the test and key basis respectively. Their Shannon entropies are denoted $H(\vec{e}_t)$ and $H(\vec{e}_k)$.} {In cases with $d>k$ subspace post-selection was used and TCS denotes the total coincidences per second. The difference between TCS and TSCS is the average number of coincidence events that were discarded due to subspace post selection per second. Additionally, error vectors, their entropies and key rates per coincidence (denoted BPC) are calculated for each subspace, where $S1-S4$ are subspace labels. We also list is the probability $\Pr(S1)-\Pr(S4)$ of obtaining result in subspace $S1-S4$ conditioned on a subspace post-selected event. Together with BPC these are used to calculate BPSC.}  {Last but not least, in $d=8,k=8$ case $F_+$ is the fidelity to the maximally entangled state we measured, which is used to lower bound the entropy $H(\vec{e}_t)$ as described in Appendix \ref{d8keyrate}. All values are rounded to the third decimal number.}
\begin{table*}[ht]
\caption{Experimental quantities $d=8,k=8$}
\begin{center}
\begin{tabular}{c|c|c|c|c}
  \toprule
    \multicolumn{5}{c} { $\boldsymbol{d=8, k=8}$ } \\
   \hline\hline

       NOISE &  0
 & 11.44
 & 33.915
 & 74.755

  \\
  \hline
          $p$ &  0
 & 0.025
 & 0.075
 & 0.15

  \\\hline
 BPSC  &$\ 2.523
 \pm 0.012 \ $ & $\ 2.183
 \pm 0.011 \ $ & $\ 1.600
 \pm 0.008 \ $ & $\ 0.830 
 \pm 0.009 \ $    \\
  \hline
  TSCS
& 3291.88
&3383.4
&3563.2
&3889.92
\\
\hline
   BPS & $\ 8307
 \pm 40 \ $ & $\ 7386
 \pm 37 \ $ & $\ 5702
 \pm 29 \ $ & $\ 3230
 \pm 35
 \ $   \\
  \hline
$F_+$ & $0.964$ & $0.943$ & $0.894$ & $0.824$ \\ \hline
$\vec{e}_k$ & 
\makecell{\tiny$(0.986,0.003,0.003,0.001,$\\\tiny$0.001,0.001,0.003,0.002)$}& \makecell{\tiny$(0.960,0.007,0.006,0.005,$\\\tiny$0.004,0.005,0.007,0.006)$}& \makecell{\tiny$(0.921,0.012,0.012,0.011,$\\\tiny$0.010,0.010,0.012,0.012)$}& \makecell{\tiny$(0.856,0.021,0.022,0.020,$\\\tiny$0.019,0.020,0.021,0.021)$}\\\hline
${H}(\vec{e}_t)$&$0.323$&$0.474$&$0.787$&$1.166$\\\hline
$H(\vec{e}_k)$&$0.153$&$0.343$&$0.613$&$1.004$\\\hline\hline
\end{tabular}
\end{center}

\label{table:fidelity}
\end{table*}

\begin{table*}[ht]
\caption{Experimental quantities $d=8,k=4\tiny$}
\begin{center}
\begin{tabular}{c|c|c|c|c|c|c}
  \toprule
  \multicolumn{7}{c} { $\boldsymbol{d=8, k=4}$ } \\
  \hline\hline
       NOISE & \multirow{4}{*}{} & 0
 & 11.45
 & 33.735
 & 73.1575
 & 164.735
  \\
     \hline
             $p$ & &  0
 & 0.025
 & 0.075
 & 0.15
& 0.3
  \\\hline
 BPSC & &$\ 1.437
 \pm 0.006 \ $ & $\ 1.301
 \pm 0.007 \ $ & $\ 1.108
 \pm 0.006 \ $ & $\ 0.789
 \pm 0.005 \ $ & $\ 0.244
  \pm 0.005  $  \\
  \hline
  TSCS &
  &3240.08
&3290.12
&3380.52
&3520.92
&3896\\\hline
BPS & &$4656
\pm19$ & $4280
\pm23$ & $3745
\pm20$ & $2777
\pm18$ & $952
\pm19$\\\hline
TCS    &              & $3250.92  $                 & $3342.6$                    & $3519.36 $                  & $3823.4$                    & $4572.84$    \\  \hline
\multirow{6}{*}{S1} & $\vec{e}_t$                 & \tiny$(0.988,0.004,0.005,0.003)$ & \tiny$(0.978,0.007,0.009,0.006)$ & \tiny$(0.959,0.013,0.014,0.014) $& \tiny$(0.925,0.024,0.026,0.025)$ & \tiny$(0.860,0.047,0.047,0.046)$ \\\cline{2-7}
                            & $\vec{e}_k$                 & \tiny$(0.935,0.020,0.027,0.018)$ & \tiny$(0.925,0.023,0.031,0.021)$ & \tiny$(0.906,0.030,0.035,0.029)$ & \tiny$(0.875,0.040,0.047,0.038)$ & \tiny$(0.815,0.060,0.065,0.060)$ \\\cline{2-7}
                            & $H(\vec{e}_t)$              & 0.109              & 0.189               & 0.307               & 0.478             & 0.806               \\\cline{2-7}
                            & $H(\vec{e}_k) $             & 0.450               & 0.505               & 0.599               & 0.701                & 0.981               \\\cline{2-7}
                            & BPC             & 1.441               & 1.306               & 1.094              & 0.758               & 0.213              \\\cline{2-7}
                            & $\Pr(S1)$ & 0.503               & 0.504               & 0.505               & 0.507               & 0.504               \\\hline
\multirow{6}{*}{S2} & $\vec{e}_t$                 & \tiny$(0.987,0.004,0.006,0.003)$ & \tiny$(0.976,0.008,0.009,0.007)$ & \tiny$(0.961,0.012,0.015,0.012)$ & \tiny$(0.930,0.023,0.024,0.023)$ & \tiny$(0.869,0.043,0.046,0.042)$ \\\cline{2-7}
                            & $\vec{e}_k$                 & \tiny$(0.935,0.020,0.028,0.017)$ & \tiny$(0.924,0.025,0.029,0.022)$ & \tiny$(0.909,0.030,0.035,0.026)$ & \tiny$(0.884,0.037,0.046,0.033)$ & \tiny$(0.823,0.057,0.064,0.056)$ \\\cline{2-7}
                            & $H(\vec{e}_t)$              & 0.117               & 0.198               & 0.296               & 0.503               & 0.768               \\\cline{2-7}
                            & $H(\vec{e}_k)$              & 0.450             & 0.506               & 0.582              & 0.739              & 0.956                \\\cline{2-7}
                            & BPC             & 1.434             & 1.295              & 1.122               & 0.820               & 0.276               \\\cline{2-7}
                            & $\Pr{(S2)}$ & 0.497                & 0.495               & 0.495               & 0.493              & 0.496               \\
  \hline\hline
\end{tabular}
\end{center}
\label{table:datad8k4}
\end{table*}

\begin{table*}[ht]
\caption{Experimental quantities $d=8,k=4$}
\begin{center}
\begin{tabular}{c|c|c|c|c|c|c}
  \toprule
   
    \multicolumn{7}{c}{ $\boldsymbol{d=8, k=2}$ } \\
  \hline\hline
       NOISE & &0
 & 11.995
 & 34.065
 & 73.3725
 & 163.845
  \\
     \hline
             $p$ & & 0
 & 0.025
 & 0.075
 & 0.15
& 0.3
  \\\hline
  BPSC & &$0.778
\pm0.005$ & $0.748
\pm0.004$ & $0.669
\pm0.005$ & $0.556
\pm0.005$ & $0.348
\pm0.004$  \\
  \hline
  TSCS & &
  3169.48
&3215.8
&3260.48
&3329.8
&3486.2\\\hline
 BPS & &$2467
\pm16$ & $2407
\pm13$ & $2183
\pm16$ & $1851
\pm17$ & $1213
\pm14$  \\
  \hline
 TCS   &               & 3233.12       & 3342.28       & 3517.6        & 3837.84       & 4540.72       \\\hline
\multirow{6}{*}{S1}  & $\vec{e}_t$                  & (0.996,0.004) & (0.993,0.007) & (0.985,0.015) & (0.974,0.026) & (0.946,0.054) \\\cline{2-7}
                     & $\vec{e}_k$                  & (0.970,0.030) & (0.970,0.030) & (0.959,0.041) & (0.950,0.050) & (0.926,0.074) \\\cline{2-7}
                     & $H(\vec{e}_t) $              & 0.037   & 0.063  & 0.111   & 0.171   & 0.302   \\\cline{2-7}
                     & $H(\vec{e}_k)$               & 0.195   & 0.192   & 0.245   & 0.286   & 0.382   \\\cline{2-7}
                     & BPC            & 0.768   & 0.744   & 0.644   & 0.543   & 0.31619027  \\\cline{2-7}
                     & $\Pr(S1)$ & 0.252   & 0.253   & 0.252   & 0.251   & 0.251   \\\hline
\multirow{6}{*}{S2}  & $\vec{e}_t$                  & (0.995,0.005) & (0.994,0.006) & (0.985,0.015) & (0.976,0.024) & (0.949,0.051) \\\cline{2-7}
                     & $\vec{e}_k$                  & (0.971,0.029) & (0.968,0.032) & (0.961,0.039) & (0.951,0.049) & (0.924,0.076) \\\cline{2-7}
                     & $H(\vec{e}_t)$               & 0.038   & 0.055   & 0.110  & 0.166   & 0.289    \\\cline{2-7}
                     & $H(\vec{e}_k) $              & 0.189  & 0.205   & 0.236  & 0.282  & 0.389  \\\cline{2-7}
                     & BPC            & 0.773   & 0.739  & 0.654   & 0.552   & 0.322  \\\cline{2-7}
                     & $\Pr(S2)$ & 0.258   & 0.252   & 0.254   & 0.257   & 0.258  \\\hline
\multirow{6}{*}{S3}  & $\vec{e}_t$                  & (0.997,0.003) & (0.994,0.006) & (0.987,0.013) & (0.974,0.026) & (0.948,0.052) \\\cline{2-7}
                     & $\vec{e}_k$                  & (0.973,0.027) & (0.971,0.029) & (0.967,0.033) & (0.953,0.047) & (0.397,0.063) \\\cline{2-7}
                     & $H(\vec{e}_t)$               & 0.032  & 0.055   & 0.101  & 0.174   & 0.295  \\\cline{2-7}
                     & $H(\vec{e}_k)$               & 0.181   & 0.189   & 0.211   & 0.272  & 0.340   \\\cline{2-7}
                     & BPC            & 0.788   & 0.756  & 0.688   & 0.555   & 0.365    \\\cline{2-7}
                     & $\Pr(S3)$ & 0.242   & 0.246   & 0.242   & 0.243   & 0.246   \\\hline
\multirow{6}{*}{S4}  & $\vec{e}_t$                  & (0.996,0.004) & (0.994,0.006) & (0.988,0.012) & (0.979,0.021) & (0.955,0.045) \\\cline{2-7}
                     & $\vec{e}_k$                  & (0.973,0.027) & (0.971,0.029) & (0.967,0.033) & (0.952,0.048) & (0.936,0.064) \\\cline{2-7}
                     & $H(\vec{e}_t) $              & 0.036    & 0.055   & 0.096   & 0.148   & 0.265   \\\cline{2-7}
                     & $H(\vec{e}_k)$               & 0.179  & 0.190   & 0.211   & 0.278   & 0.344   \\\cline{2-7}
                     & BPC             & 0.785  & 0.755   & 0.693   & 0.574   & 0.391   \\\cline{2-7}
                     & $\Pr(S4)$ & 0.249   & 0.249   & 0.251   & 0.249   & 0.245 \\\hline\hline
\end{tabular}
\end{center}
\label{table:datad8k2}
\end{table*}

\begin{table*}[ht]
\caption{Experimental quantities $d=4,k=4$}
\begin{center}
\begin{tabular}{c|c|c|c|c}
  \toprule
    \multicolumn{5}{c} { $\boldsymbol{d=4, k=4}$ } \\
   \hline
    \hline
 NOISE & 0
 & 10.825
 & 34.47
 & 73.615
   \\
     \hline
             $p$ &  0
 & 0.025
 & 0.075
 & 0.15

  \\\hline
 BPSC & $1.420
\pm0.01$ & $1.195
\pm0.011$ & $0.844
\pm0.008$ & $0.365
\pm0.008$
  \\
  \hline
  TSCS & 1633.4
&1675.36
&1770.68
&1925.56\\
\hline
  BPS & $2319
\pm16$ & $2002
\pm18$ & $1495
\pm14$ & $703
\pm15$ \\
  \hline
 $\vec{e}_t$    & (0.988,0.004,0.005,0.003) & (0.968,0.011,0.011,0.010) & (0.934,0.021,0.023,0.022) & (0.880,0.039,0.041,0.040) \\\hline
$\vec{e}_k$    & (0.931,0.020,0.028,0.021) & (0.916,0.026,0.030,0.028) & (0.883,0.035,0.044,0.038) & (0.832,0.054,0.057,0.057) \\\hline
$H(\vec{e}_t)$ & 0.112              & 0.253             & 0.454                 & 0.719                \\\hline
$H(\vec{e}_k)$ & 0.468               & 0.549               & 0.703              & 0.918  \\\hline\hline
\end{tabular}
\end{center}

\label{table:d4k4data}
\end{table*}

\begin{table*}[ht]
\caption{Experimental quantities $d=4,k=2$}
\begin{center}
\begin{tabular}{c|c|c|c|c|c}
  \toprule
     \multicolumn{6}{c}{ $\boldsymbol{d=4, k=2}$ } \\
      \hline
     NOISE & &0
 & 12.855
 & 33.16
 & 69.485
   \\
      \hline
              $p$ & & 0
 & 0.025
 & 0.075
 & 0.15

  \\\hline
   BPSC & &$0.771
\pm0.006$ & $0.684
\pm0.006$ & $0.571
\pm0.005$ & $0.384
\pm0.005$  \\
  \hline
  TSCS & &1610.56 
  & 1647.64
&1677.68
&1736.48\\
\hline
 BPS & &$1242
\pm10$ & $1126
\pm10$ & $957
\pm8$ & $667
\pm9$   \\
 \hline
TCS &                & 1640.6        & 1703.92       & 1775.56       & 1912.88       \\\hline
\multirow{6}{*}{S1}     & $\vec{e}_t$    & (0.995,0.005) & (0.989,0.011) & (0.975,0.025) & (0.956,0.044) \\\cline{2-6}
                        & $\vec{e}_k$    & (0.969,0.021) & (0.964,0.036) & (0.951,0.049) & (0.929,0.071) \\\cline{2-6}
                        & $H(\vec{e}_t)$ & 0.045   & 0.090   & 0.167   & 0.261   \\\cline{2-6}
                        & $H(\vec{e}_k)$ & 0.198   & 0.224   & 0.282   & 0.370   \\\cline{2-6}
                        & BPC            & 0.758   & 0.686   & 0.550   & 0.369   \\\cline{2-6}
                        & $\Pr(S1)$      & 0.496         & 0.498         & 0.491         & 0.497         \\\hline
\multirow{6}{*}{S2}     & $\vec{e}_t$    & (0.996,0.004) & (0.988,0.012) & (0.978,0.021) & (0.956,0.044) \\\cline{2-6}
                        & $\vec{e}_k$    & (0.973,0.027) & (0.963,0.037) & (0.957,0.043) & (0.936,0.064) \\\cline{2-6}
                        & $H(\vec{e}_t)$ & 0.035   & 0.091   & 0.152   & 0.258   \\\cline{2-6}
                        & $H(\vec{e}_k)$ & 0.181  & 0.228  & 0.258    & 0.343   \\\cline{2-6}
                        & BPC            & 0.785   & 0.681   & 0.590   & 0.399  \\\cline{2-6}
                        & $\Pr(S2)$      & 0.504         & 0.502         & 0.509         & 0.503      \\\hline\hline  
 \end{tabular}
\end{center}

\label{table:d4k2data}
\end{table*}

 \begin{table*}[ht]
\caption{Experimental quantities $d=2,k=2$}
\begin{center}
\begin{tabular}{c|c|c|c|c}
  \toprule
     \multicolumn{5}{c}{ $\boldsymbol{d=2, k=2}$ } \\
      \hline
     NOISE & 0
 & 9.26
 & 33.17
 & 65.57
   \\
      \hline
              $p$ &  0
 & 0.025
 & 0.075
 & 0.15

  \\\hline
  BPSC & $0.805
\pm0.012$ & $0.669
\pm0.012$ & $0.424
\pm0.012$ & $0.166
\pm0.011$
  \\
  \hline
  TSCS 
  &816.44
  &832.4
&882.52
&935.6
\\\hline
  BPS & $657
\pm10$ & $557
\pm10$ & $374
\pm11$ & $156
\pm10$
  \\\hline
$\vec{e}_t$    & (0.995, 0.005) & (0.983,0.017) & (0.958,0.042) & (0.924,0.076)  \\\hline
$\vec{e}_k$    & (0.978,0.022)  & (0.968,0.032) & (0.940,0.060) & (0.907, 0.093) \\\hline
$H(\vec{e}_t)$ & 0.045 & 0.124   & 0.250   & 0.387    \\\hline
$H(\vec{e}_k)$ & 0.150    & 0.206   & 0.327   & 0.447   \\\hline\hline
\end{tabular}
\end{center}

\label{table:d2k2data}
\end{table*}


\end{document}